\documentclass{desyproc}

\begin{document}
\title{MADMAX: A new Dark Matter Axion Search using a Dielectric Haloscope}

\author{{\slshape  
B\'ela Majorovits$^{1}$ and Javier Redondo$^{2}$ on behalf of the MADMAX Working Group\\
(A.\,Caldwell$^{1}$, G.\,Dvali$^{1,3}$, 
C.\,Gooch$^{1}$, A.\,Hambarzumjan$^{1}$,  B.\,Majorovits$^{1}$, A.\,Millar$^{1}$, 
G.\,Raffelt$^{1}$, J.\,Redondo$^{2}$, O.\,Reimann$^{1}$, F.\,Simon$^{1}$, F.\,Steffen$^{1}$)
}\\[1ex]
$^1$Max-Planck-Institut f\"ur Physik, M\"unchen, Germany,\\
$^2$University of Zaragoza, Spain,\\
$^3$Ludwig Maximilians Universit\"at M\"unchen, Germany,
}

\contribID{familyname\_firstname}

\confID{13889}  
\desyproc{DESY-PROC-2016-03}
\acronym{Patras 2016} 
\doi  

\maketitle

\begin{abstract}
The axion is an intriguing dark matter candidate emerging from the Peccei--Quinn solution to the strong CP problem.  
Current experimental searches for axion dark matter focus on the axion mass range below 40~$\mu$eV. 
However, if the Peccei--Quinn symmetry is restored after inflation 
the observed dark matter density
points to an axion mass around 100~$\mu$eV.
A new project based on axion-photon conversion at the transition between different dielectric media is presented. 
By using $\sim 80$ dielectric discs,
the emitted power could be enhanced by a factor of $\sim 10^5$ over that from a single mirror (flat dish antenna).
Within a 10 T magnetic field, this could be enough to detect $\sim 100$\,$\mu$eV axions with HEMT linear amplifiers.
The design for an experiment is proposed.
Results from noise, 
transmissivity and reflectivity measurements obtained in a prototype setup are presented. 
The expected sensitivity is shown.
\end{abstract}

\section{Introduction}
The axion is a hypothetical low-mass boson which emerges 
as a consequence of the Peccei--Quinn (PQ) mechanism, which explains the absence of CP-violating effects in quantum chromodynamics. 
%
%
Axions are attractive candidates for cold dark matter (DM) as they are produced non-thermally in the early universe. 
There are two classes of axion DM scenarios:
the PQ symmetry is broken before inflation and either
(i)~never restored thereafter or (ii)~restored during reheating, subsequently breaking again.
%
%
In scenario~(i), axion DM is produced through the re-alignment mechanism with a dependence on one common initial misalignment angle $\theta_\mathrm{I}$ in the observable universe. A wide axion mass range can be made to agree with the observed DM density as $\theta_\mathrm{I}$ is a free parameter, though values around $m_a\sim 10~\mu$eV corresponding to $\theta_\mathrm{I} \sim$\,1 are considered to be most natural. 
In scenario~(ii), the observable universe contains many patches with different $\theta_\mathrm{I}$ values, whose average fixes the axion DM contribution from the re-alignment mechanism.
With additional contributions from decays of axion strings and domain walls, 
an agreement with the DM density is found in the `high-mass' region around $m_a\sim 100~\mu$eV~\cite{Hiramatsu}.

Most efforts for detecting DM axions, notably the
ADMX and ADMX-HF experiments \cite{Rybka:2013,Brubaker:2016ktl}, focus
on $m_a\lesssim 40$~$\mu$eV. These experiments rely on 
the resonant enhancement by a cavity to boost the axion-photon conversion 
in a magnetic field and so gain sensitivity to axion DM. 
As resonant cavities are most effective for $m_a$ of ${\cal O}(10)~\mu$eV,
a cavity-based search for axion DM in the 100~$\mu$eV range
seems to be highly challenging.
New methods are needed to cover this 100~$\mu$eV region.
%



\begin{figure}[t!]
\centerline{\includegraphics[width=0.4\textwidth]{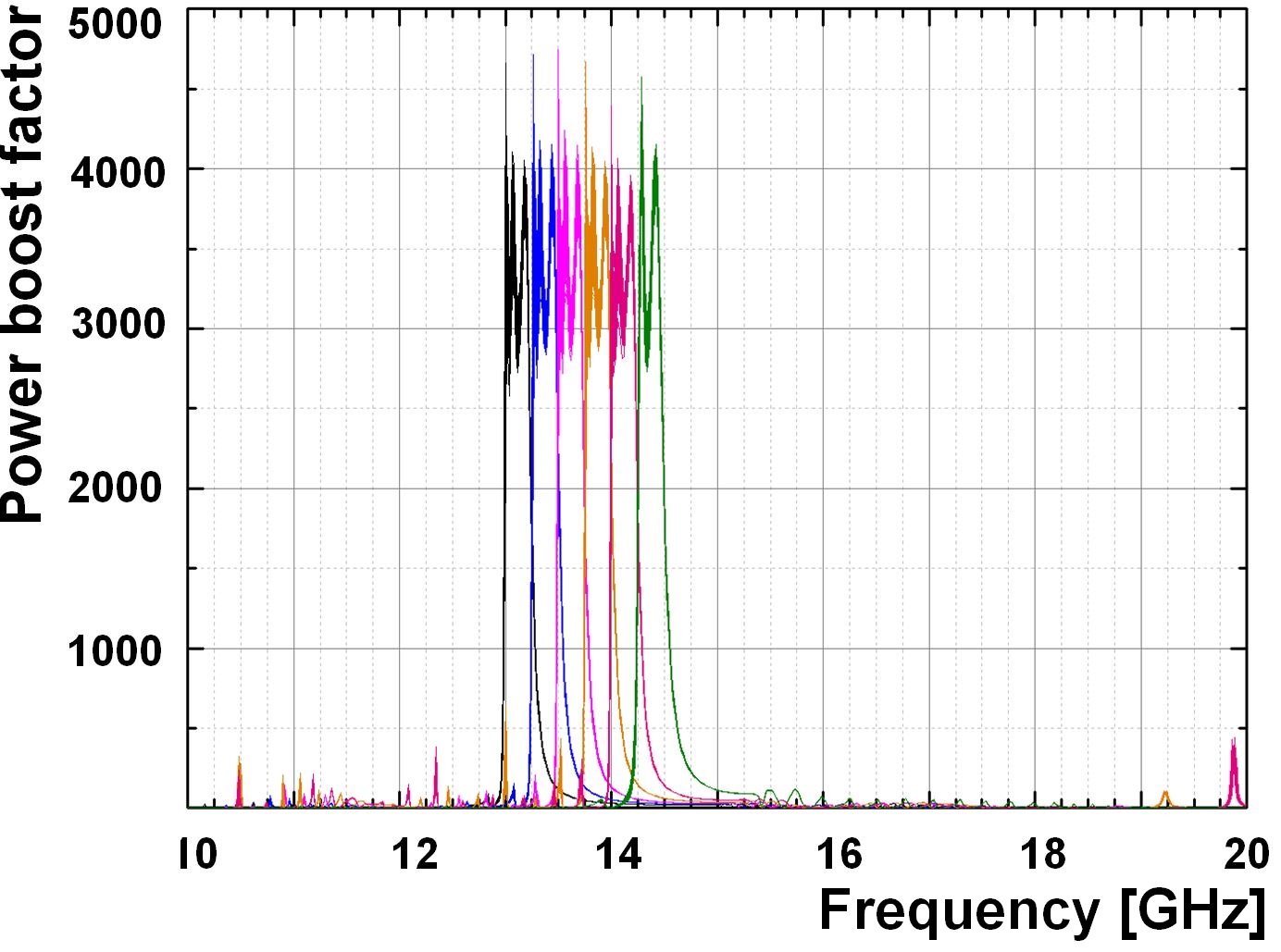}
\includegraphics[width=0.45\textwidth]{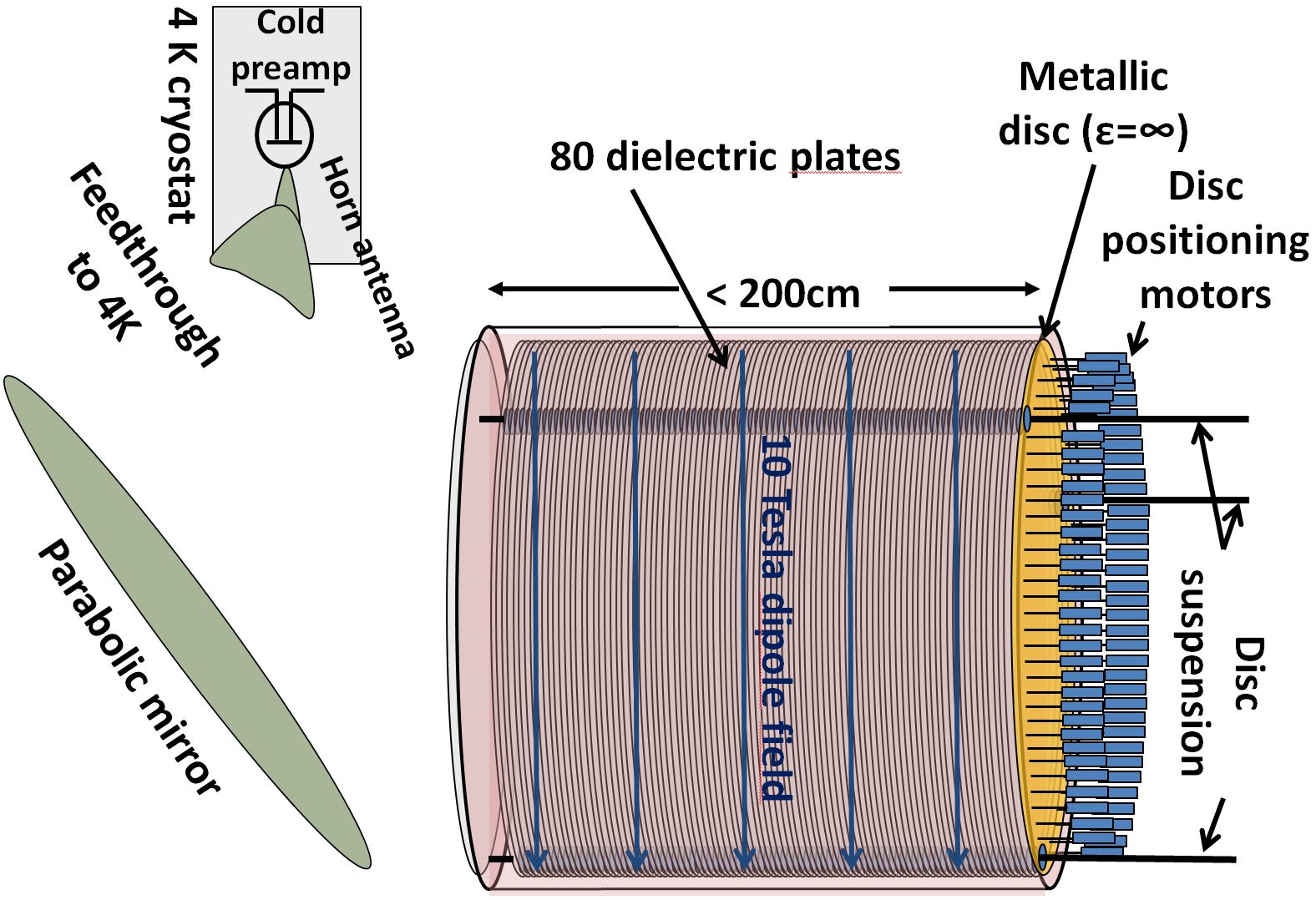}}
\caption{Left: The power boost factor in a dielectric haloscope with $N=20$ LaAlO$_3$ discs. 
Right: Conceptual idea of a dielectric haloscope to search for DM axions with $m_a\sim 100~\mu$eV.}\label{Fig:boost}
\end{figure}

\section{Dielectric haloscope concept and experimental idea}
To scan the $m_a \gtrsim 40~\mu$eV range, we exploit a novel dielectric haloscope
approach
derived from the concepts described in~\cite{Horns:2013,Jaeckel:2013}. 
The expected axion-induced radiation
power density 
at a transition between materials
with different dielectric constant $\epsilon$ in a magnetic field is
\begin{equation}
P/A = 2.2\times 10^{-27}~({\rm W/m}^{2})\,(B_{||}/10\,{\rm T})^{2}
\, C_{a\gamma}^{2} \, F(\Delta_{\epsilon_{1},\epsilon_{2}}),
\label{eq:AxionInducedPower}
\end{equation}
where $A$ is the surface area,
$B_{||}$ the magnetic field parallel to the surface, and $C_{a\gamma}$ an ${\cal O}(1)$ constant
quantifying the model-dependence of the axion-photon
coupling $|g_{a\gamma}|=\alpha |C_{a\gamma}|/(2\pi f_{a}$)~\cite{PDG}, 
with the fine-structure constant $\alpha$ and the PQ breaking scale $f_a$.
$F(\Delta_{\epsilon_{1},\epsilon_{2}})$ is a function depending on the
dielectric constants of the two media. For a metallic mirror in vacuum
$F(\Delta_{0,\infty})=1$~\cite{Horns:2013}. 

\begin{wrapfigure}{r}{0.5\textwidth}
\centerline{
\includegraphics[width=0.46\textwidth]{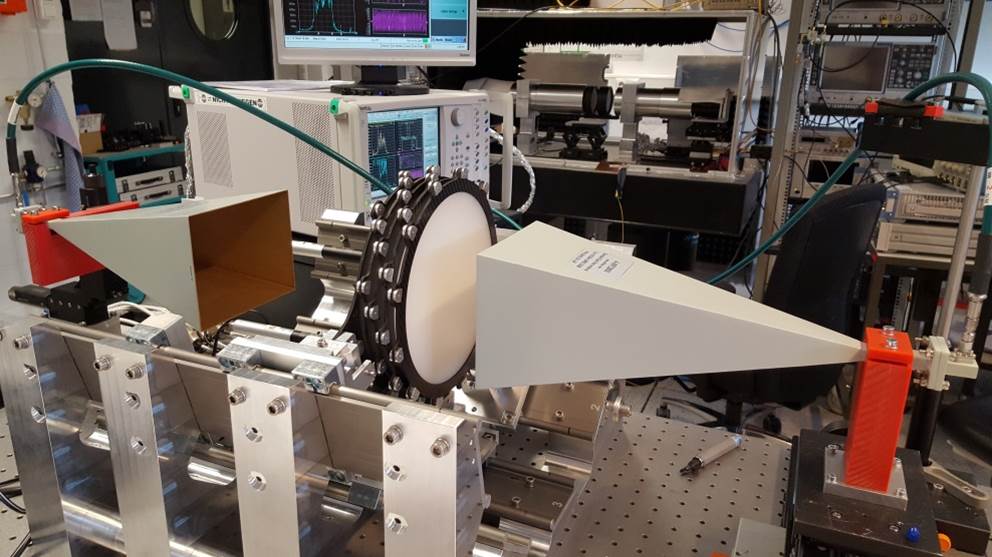}
}
\caption{
Preliminary prototype setup for transmissivity measurements with 
four 
Al$_2$O$_3$ discs.
}
\label{Fig:seed_setup}
\end{wrapfigure}
To obtain a detectable power of $P\sim 10^{-23}$\,W,
one would need
a mirror with an area of $A\sim 5000$ m$^2$ magnetized with a 10~T field parallel to its surface, 
which is not possible with current technology.

Instead, we propose building a dielectric haloscope: 
a series of $N$ high-$\epsilon$ material discs with $A\sim 1$~m$^2$ placed 
in front of a metallic mirror of the same area, contained in a $\sim$~10~T magnetic field. 
Unlike a traditional resonant cavity, this device is open on one side to facilitate broadband searches. 
One can adjust the spacings between the discs to achieve constructive interference, leading to
to a significant boost of the expected axion-induced power $P$ 
over a sizeable bandwidth. 
The power boost factor is defined as the
power generated in the multi-disc dielectric haloscope normalized to that generated
by a single metallic mirror with the same area $A\sim 1$~m$^2$. 
This factor depends on the frequency $\nu_a=m_a/2\pi$, the number of discs $N$, 
their spacings and their $\epsilon$ values.

Figure~\ref{Fig:boost}, left, shows the power boost factor as a function of $\nu_a$, 
generated by electromagnetic (EM) simulations for a setup consisting of $N=20$~discs made from LaAlO$_3$ ($\epsilon \sim 24$):
factors~$\gtrsim 3\times 10^{3}$ within a bandwidth of 250~MHz can be achieved.
%
%
Moreover, the frequency range with a sizeable power boost
can be seamlessly shifted by changing the spacings between the discs.

Further investigations using EM modelling 
show that
the 
area under the power
boost factor curve as a function of frequency $\nu_a$
scales linearly with the number of discs $N$. 
Based on this area law, power boost factors of  $\sim 10^{5}$ over a 40\,MHz wide frequency range
seem feasible with 
$N\sim 80$ discs made from LaAlO$_{3}$.
Using discs with a diameter of 1~m in a 10~T dipole field, a total axion-induced
power of $\sim 10^{-23}$~W could then be generated in the high-mass region around $m_a\sim 100~\mu$eV. 
A sketch of the proposed MADMAX setup is shown in Fig.~\ref{Fig:boost}, right.

While this is a mechanically challenging setup, it has the advantage of flexibility.
One can adopt a significantly broadband search strategy, 
probing a large frequency range within a reasonable measurement time.
Once the broad bandwidth search leads to evidence for a signal,
the  dielectric haloscope can be tuned to a higher boost factor across a narrower bandwidth.
This would enhance the signal-to-noise ratio, allowing the fast confirmation of a detection.

One of the big challenges in the experiment will be the availability of a $\sim$~10 T 
dipole magnet that allows to house $\sim$~1~m diameter discs over
a length of up to 2~m.
Presently two design concepts are under discussion, the canted cosine theta \cite{canted}
and the racetrack \cite{racetrack} designs. Initial investigations indicate that both design concepts are suitable.
In the near future we will see which of these is better suited for the current 
experimental proposal.

\section{First measurements and sensitivity projection}
\begin{wrapfigure}{r}{0.5\textwidth}
\centerline{
\includegraphics[width=0.45\textwidth]{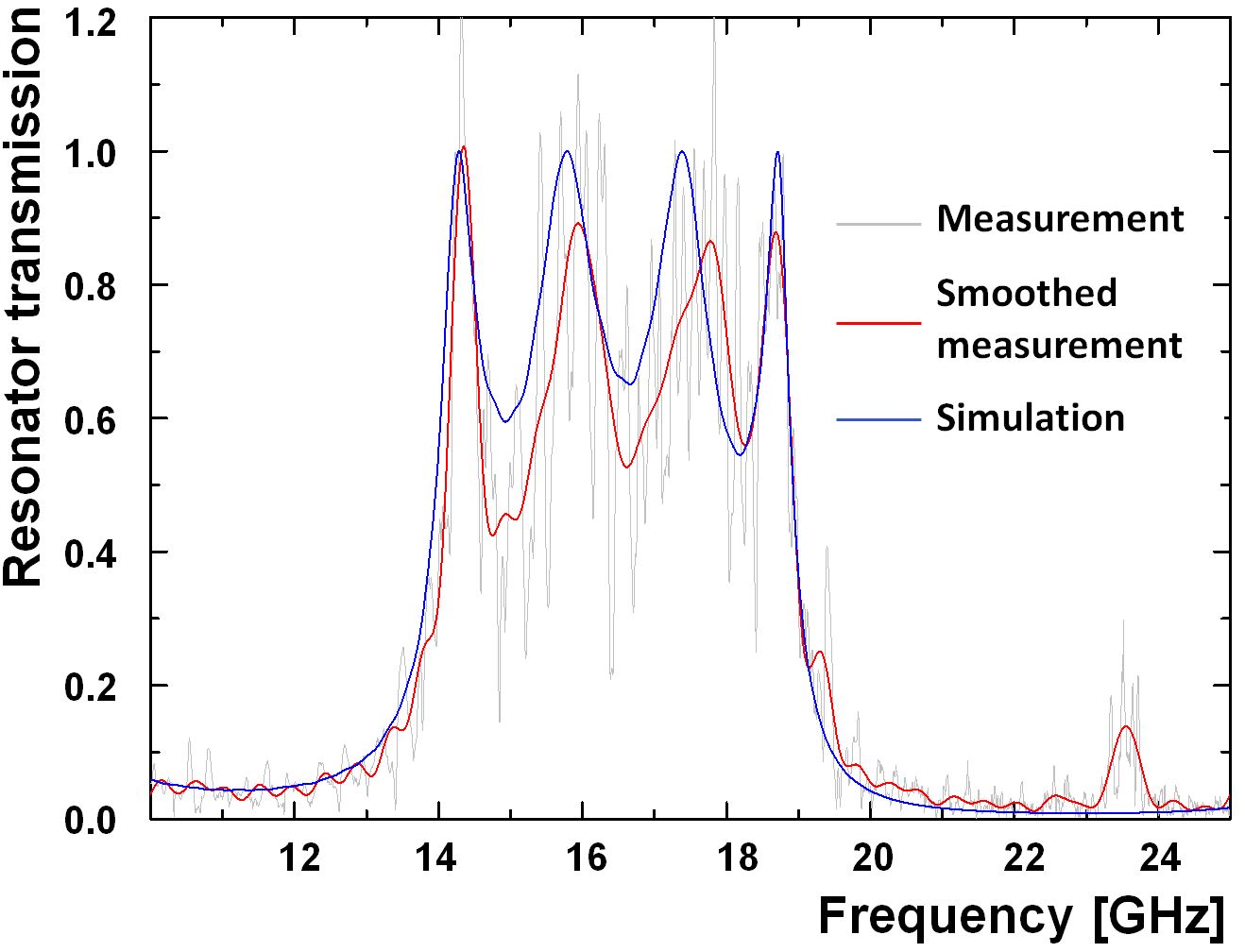}
}
\caption{First transmission measurement (gray) taken with five Al$_2$O$_3$ 
discs between receiver and emitter. The smoothed measurement curve
 (red) can be compared to the behaviour predicted by simulation (blue). 
}\label{Fig:cavity_measurements}
\end{wrapfigure}
To determine the required signal strength, a radiometer based on heterodyne detection using a 
HEMT preamplifier \cite{hemt_datasheet} has been built. It was used for a first 
measurement of a weak signal at 20\,GHz with $3\times10^{-21}$~W  at room temperature. 
The fake signal was detected within one week measurement time with
6\,$\sigma$ separation from background. 
According to the data sheet, operating the preamplifier at 4\,K temperature 
reduces its noise by another two orders of magnitude allowing for the detection of signals with power of the order of
$3\times10^{-23}$~W, if the noise contribution of the haloscope
itself is low enough.

As transmissivity and reflectivity of the haloscope
are correlated with the boost factor curve, they
can be used to verify the simulated boost factor behaviour.
We use this to test our calculations of the boost factor
and to potentially aid correct disc placement.

This has been done in the prototype setup shown in Fig.~\ref{Fig:seed_setup}
consisting of five sapphire (Al$_2$O$_3$ with $\epsilon \sim 10$) discs 
with 200\,mm diameter each. Discs are positioned by 
precision motors with a precision of roughly 15\,$\mu$m. 
The uncertainty on positioning is due to the mechanical setup rather than the limitations of the motors.
A comparison of a transmissivity measurement with simulation is shown in
Fig.~\ref{Fig:cavity_measurements}. While details of the 
transmissivity behaviour slightly differ, the general behaviour 
is encouragingly similar, especially the position of the peak at highest frequencies.
The same has been verified for reflectivity measurements
using group delay peaks.

\begin{wrapfigure}{r}{0.5\textwidth}
\includegraphics[width=0.5\textwidth]{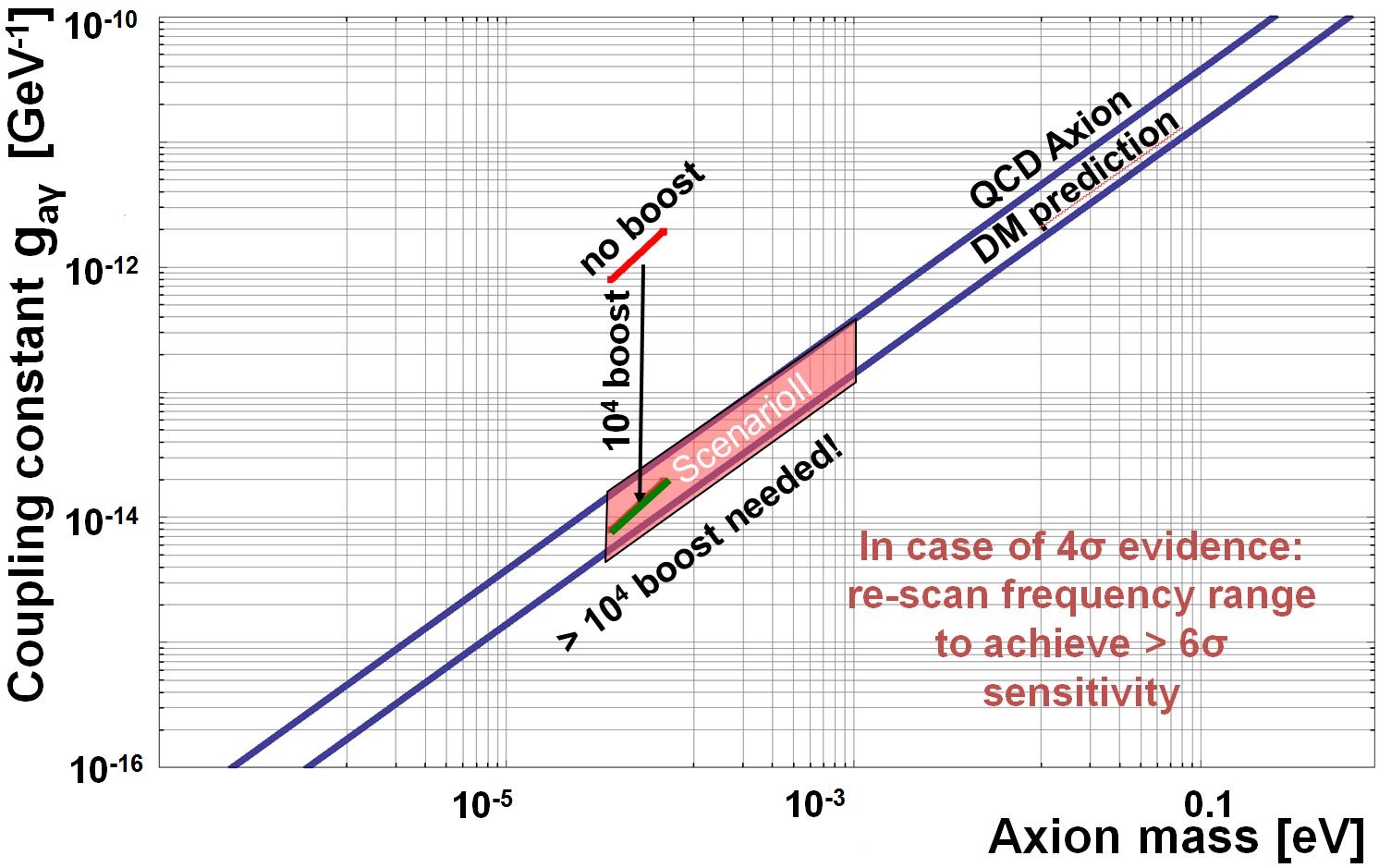}
\caption{Projected MADMAX sensitivity}\label{Fig:sensitivity}
\label{sec:figures}
\end{wrapfigure}

A long term reflectivity measurement using four sapphire discs and a metallic 
mirror was performed to monitor the 
stability of the group delay peak.
The variations of the peak position, attributed to mechanical (vibration) or thermal 
(contraction or expansion) changes in the setup, are of the order of 1~MHz, 
substantially narrower than the envisioned bandwidth of the boost factor. 
Long-term measurements can be performed over a sizeable bandwidth even under these experimental conditions (without vibrational damping and precise 
climate control).

To estimate the sensitivity of the proposed axion DM search experiment, we assume a setup consisting of 
$N=80$ discs with $A=1$~m$^2$ surface area each, achieving a power boost factor of $\gtrsim$~10$^4$ 
over a bandwidth of $50$ MHz, and cryogenic preamp with a noise level of 8~K. 
Inside  a 10~T dipole magnetic field this would allow for a 
scanning of the well-motivated QCD axion DM parameter space around $m_a\sim100~\mu$eV, 
assuming purely axionic DM, within a few years. 
The projected sensitivity  is depicted in Fig.~\ref{Fig:sensitivity}.

\section{Outlook and Conclusions}

Axions are amongst the best motivated  DM candidates, with interesting and unique phenomenology. 
However, `high mass'  axions around $100~\mu$eV
that explain the observed DM density in scenarios in which PQ symmetry breaking occurs after inflation are beyond the reach of current experiments. A novel approach to study these axions is a dielectric haloscope that consists of a series of dielectric discs placed in a magnetic field,
which can be positioned to enhance the axion-induced EM power to a detectable level.
%
%
A comparison of simulations with first measurements indicate that this approach could scan the interesting QCD axion DM mass range within a reasonable time span.

\vspace{0.1cm}

{\bf  Acknowledgments:}
The prototype setup used for first measurements was partly funded as a seed project by the DFG 
Excellence Cluster Universe (Grant No.\ EXC 153).



\begin{footnotesize}

\end{footnotesize}


\end{document}